\documentclass[iop,revtex4]{emulateapj}

\begin{document}

\renewcommand{\topfraction}{1.0}
\renewcommand{\bottomfraction}{1.0}
\renewcommand{\textfraction}{0.0}

\title{Spectroscopic orbits of subsystems in  multiple stars. III.}

\author{Andrei Tokovinin}
\affil{Cerro Tololo Inter-American Observatory, Casilla 603, La Serena, Chile}
\email{atokovinin@ctio.noao.edu}

\begin{abstract}
Spectroscopic  orbits  are computed  for  inner  pairs  in six  nearby
hierarchical  multiple   systems  (HIP  35733,   95106/95110,  105441,
105585/105569,  105947, and 109951).   Radial velocities  and resolved
measurements,  when available,  are used  to derive  combined  sets of
outer  orbital elements for  three systems.   Each multiple  system is
discussed individually.  Additionally, HIP~115087  is a simple 7.9 day
single-lined binary. Although the minimum companion mass is substellar
(in the brown dwarf desert regime),  it appears to be a 0.2 solar-mass
star     in     a     low-inclination     orbit.      \keywords{stars:
  binaries:spectroscopic}
\end{abstract}

\maketitle

\section{Introduction}
\label{sec:intro}

Formation of  close binaries is  still an unsolved  problem. Processes
that bring their components from the initial large separations to much
closer orbits and define the distributions of periods, eccentricities,
and  mass ratios  are  not fully  understood.   Statistics of  orbital
elements in large and  well-defined samples provide an essential input
for  solving this problem.   However, even  in the  solar neighborhood
many  close binaries  still lack  orbits.  This  work  contributes new
orbits  of nearby  low-mass stars,  mostly components  of hierarchical
multiple  systems.   Dynamics of  triple  systems  coupled with  tidal
friction is  one of  the processes leading  to the formation  of close
binaries.  Eventually  we hope to compare  theoretical predictions for
this  formation  channel   \citep[e.g.][]{Moe2018}  with  the  orbital
statistics in hierarchical systems.

\begin{figure}
\epsscale{1.1}
\plotone{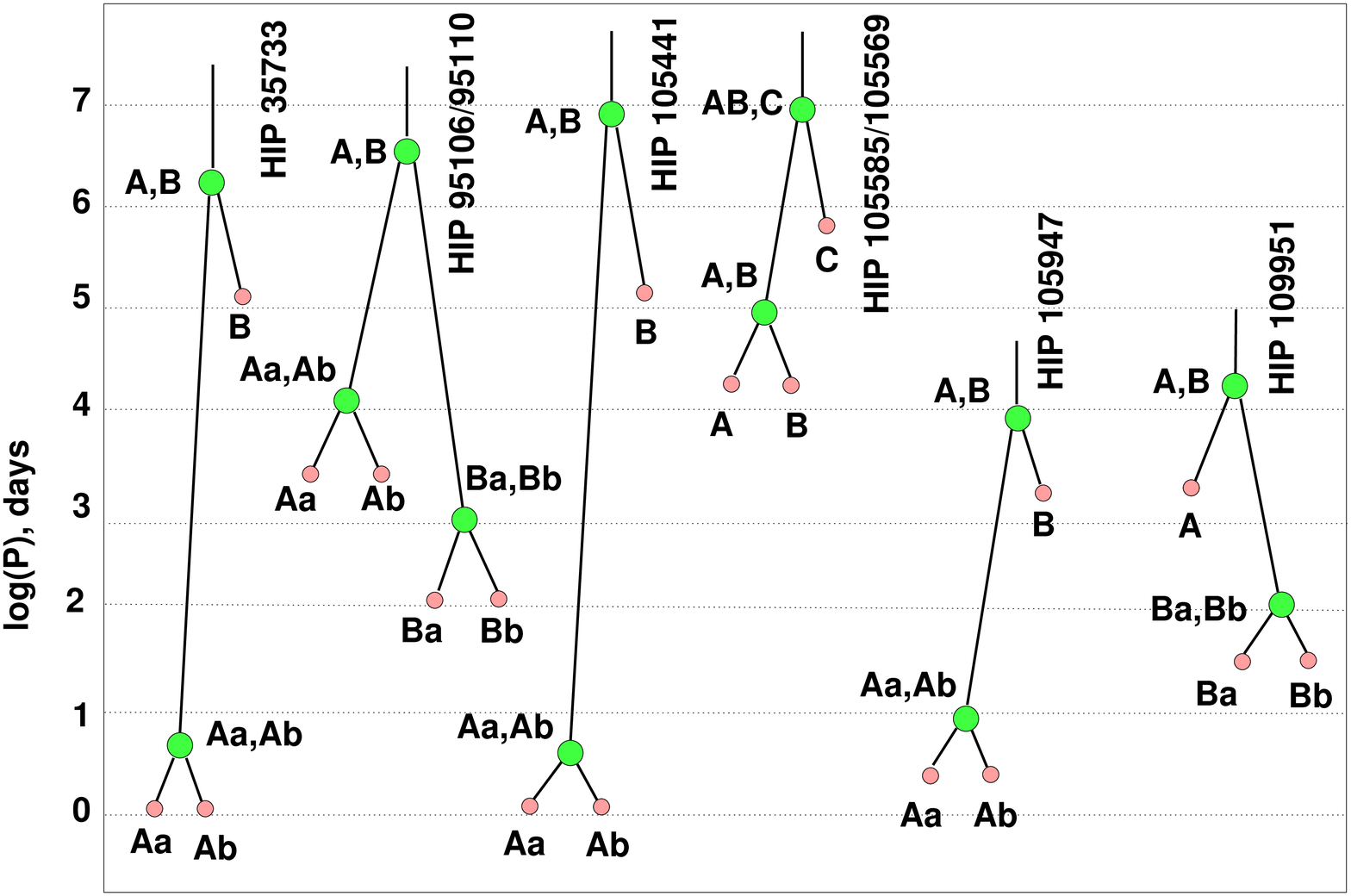}
\caption{Structure of six hierarchical systems studied here. Large
  green circles denote elementary binaries, while their vertical
  position corresponds to the logarithm of the orbital period. Small
  red circles denote individual stars.
\label{fig:mobile}
}
\end{figure}

\begin{deluxetable*}{c c rr   l cc rr  c }
\tabletypesize{\scriptsize}     
\tablecaption{Basic parameters of observed multiple systems
\label{tab:objects} }  
\tablewidth{0pt}                                   
\tablehead{                                                                     
\colhead{WDS} & 
\colhead{Comp.} &
\colhead{HIP} & 
\colhead{HD} & 
\colhead{Spectral} & 
\colhead{$V$} & 
\colhead{$V-K$} & 
\colhead{$\mu^*_\alpha$} & 
\colhead{$\mu_\delta$} & 
\colhead{$\overline{\omega}$\tablenotemark{a}} \\
\colhead{(J2000)} & 
 & &   &  
\colhead{type} & 
\colhead{(mag)} &
\colhead{(mag)} &
\multicolumn{2}{c}{ (mas yr$^{-1}$)} &
\colhead{(mas)} 
}
\startdata
07223$-$3555 &A & 35733 & 58038  &  F8    & 7.01  & 1.20 & $-$45 & +96  & 16.93  \\ 
             &B & \ldots & \ldots& F6?    & 7.87  & 1.18 & $-$39 & +98  & 16.94 \\
19209$-$3303 &A & 95106 & 181199 &  G0V    &  8.16 & 1.77 & 3     & $-$35 &  18.82   \\
             &B & 95110 & \ldots &  K2V    & 10.26 & 2.84 & 8     & $-$141 &  21.03  \\
21214$-$6655 &A & 105441& 202746 &  K2V    & 8.77  & 1.62 & 95 & $-$100  & 31.27     \\
             &B & \ldots& \ldots &  K7V    & 10.60 & 3.59 & 105 & $-$85   & 31.71   \\ 
21232$-$8703 &AB& 105585& 198477 &  G8III  & 8.10  & 1.14 & 103 & $-$88    &  \ldots  \\
             &C & 105569& \ldots &  G5     & 8.94  & 1.51 & 126 & $-$86    &  13.99  \\
21274$-$0701 &AB& 105947& 204236 &  F8/G0V & 7.52  & 1.44 & 8  & $-$95 & 15.36            \\ 
22161$-$0705 &AB& 109951& 211276 &  G5V    & 8.74  & 1.92 & 103  & $-$321 & 15.12            \\ 
23186$-$5818 &A & 115087& 219709 &  G1IV   & 7.52  & 1.52 & 214 & $-$156 & 23.35            
\enddata
\tablenotetext{a}{Proper motions and parallaxes are 
  from the {\it Gaia} DR2 \citep{Gaia}}
\end{deluxetable*}

The    multiple     systems    studied    here     are    listed    in
Table~\ref{tab:objects}.  Most  objects belong to the  67-pc sample of
solar-type  stars \citep{FG67a}.   Some spectroscopic  subsystems were
discovered in  the large survey by \citet{N04},  while others resulted
from  the   radial  velocity  (RV)  study  of   wide  visual  binaries
\citep{survey}.  Six  systems contain additional  components, i.e.  are
hierarchical;      their     structure      is      illustrated     in
Figure~\ref{fig:mobile}.   These systems are  featured in  the updated
multiple star catalog \citep{MSC}.  The large range of orbital periods
covering 7 orders  of magnitude is worth noting.   This work continues
previous publications of spectroscopic orbits based on high-resolution
spectra  taken  at the  CTIO  1.5  m telescope  \citep{paper1,paper2}.
Spectra   from  the   new  Network   of  Robotic   Telescopes  Echelle
Spectrographs (NRES) instrument, also based  at CTIO, are used here as
well.

The data and methods are briefly recalled in Section~\ref{sec:obs}, where
the new orbital elements are also given. Then in Section~\ref{sec:obj}
each system is discussed in some detail. The paper closes with a short
summary in Section~\ref{sec:sum}.

\section{Observations and data analysis}
\label{sec:obs}

\subsection{Spectroscopic observations}

Most spectra  used here were taken  with the 1.5 m  telescope sited at
the  Cerro  Tololo  Inter-American  Observatory (CTIO)  in  Chile  and
operated        by       the        SMARTS       Consortium.\footnote{
  \url{http://www.astro.yale.edu/smarts/}}  The   observing  time  was
allocated through NOAO.   Observations were made
with the CHIRON spectrograph \citep{CHIRON} by the telescope operators
in service mode.  In two runs,  2017 August and 2018 March, the author
also observed in classical mode.  Most spectra are taken in the slicer
mode with a resolution of $R=80,000$ and a signal to noise ratio of at
least 20.  Thorium-argon calibrations were recorded for each target.

In    2017    September,    the    new    NRES    fiber-fed    echelle
spectrograph\footnote{\url{https://lco.global/observatory/instruments/nres/}}
installed at the 1 m telescope of the Las Cumbres Observatory, located
at CTIO, became  available for users \citep{NRES}.  It  has a spectral
resolution of  $R=53,000$.  The  observations are scheduled  using the
web  portal and  made  robotically.   The data  are  processed by  the
pipeline.   Both the  instrument and  its  pipeline are  new and  were
offered in  the ``shared-risk'' mode. Despite this,  I obtained useful
data from NRES, extending the time coverage available with CHIRON.

For some objects, I also  use RVs measured in 2008
with the  Du Pont echelle at Las  Campanas and in 2010  with the Fiber
Echelle (FECH) at CTIO, published in \citep{LCO}.

\subsection{Radial velocities by cross-correlation}

I use here the  reduced and wavelength-calibrated spectra delivered by
the CHIRON and NRES pipelines.  The spectra were cross-correlated with
the digital binary mask (template)  based on the solar spectrum stored
in   the  NOAO  archive   \citep[see][for  more   details]{LCO}.   The
cross-correlation function (CCF) is computed  over the RV span of $\pm
200$\,km~s$^{-1}$   in   the   spectral   range   from   4500\,\AA   ~to
6500\,\AA. Portions of the CCFs  around each dip are approximated by one
or several  Gaussian curves.  After  the first iteration,  the centers
and  widths  of each  component  are  determined,  and in  the  second
iteration the fitting area is  adjusted accordingly.  I do not provide
formal errors  of RVs and of  other parameters resulting  from the CCF
fits,  as  they  are very  small  and  do  not characterize  the  real
precision of the results.  The  RV precision of CHIRON is dominated by
systematic effects  and is estimated  from residuals to the  orbits at
$\sim$0.1 km~s$^{-1}$ \citep[see][]{paper1,paper2}. 

The RVs determined by cross-correlation with the solar spectrum should
be on the absolute scale.  In  2018 March, I observed with CHIRON 4 RV
standards  from \citet{Udry98},  namely HD  73667, 82106,  125184, and
140538. The mean RV  difference (CHIRON$-$Udry) is $+0.16$ km~s$^{-1}$
and its rms scatter is 0.03\,km~s$^{-1}$.

The  NRES  pipeline delivers  spectra  with  a preliminary  wavelength
calibration.  Observations  of the single star HIP~98698  with
both NRES  and CHIRON established that  the RVs derived  from the NRES
spectra  should be  corrected by  $+3.80$\,km~s$^{-1}$ to  match those
from CHIRON. This fixed correction  is applied here to all RVs derived
from the NRES spectra.  The  preliminary NRES data products do not yet
properly  track  instrument   changes  using  simultaneous  comparison
spectrum. The RVs of telluric lines derived by the CCF method show an rms
scatter  of 0.33  km~s$^{-1}$.   For comparison,  the  rms scatter  of
telluric RVs in the  CHIRON spectra is 0.055~km~s$^{-1}$.  Considering
this  test,  the RVs  derived  here from  the  NRES  spectra may  have
systematic errors on  the order of 1\,km~s$^{-1}$ or  less.  It should
be stressed  that these results characterize the  NRES instrument and
its pipeline in their initial (not yet optimized) state.

\subsection{Speckle interferometry}

Information  on   the  resolved  subsystems  is   retrieved  from  the
Washington Double Star Catalog, WDS \citep{WDS}. It is complemented by
recent  speckle  interferometry  at  the SOAR  telescope.  The  latest
publication \citep{SAM18}  contains references to  the previous papers
of  this  series. 

\subsection{Orbit calculation}

Orbital elements and their errors were determined by the least-squares
fits with  weights inversely proportional to the  adopted  errors.
The          IDL          code          {\tt          orbit}\footnote{
  \url{http://www.ctio.noao.edu/\~{}atokovin/orbit/}} was used. It can
fit   spectroscopic,    visual,   or   combined   visual/spectroscopic
orbits. Formal  errors of orbital  elements are determined  from these
fits.

\begin{figure}
\epsscale{1.1}
\plotone{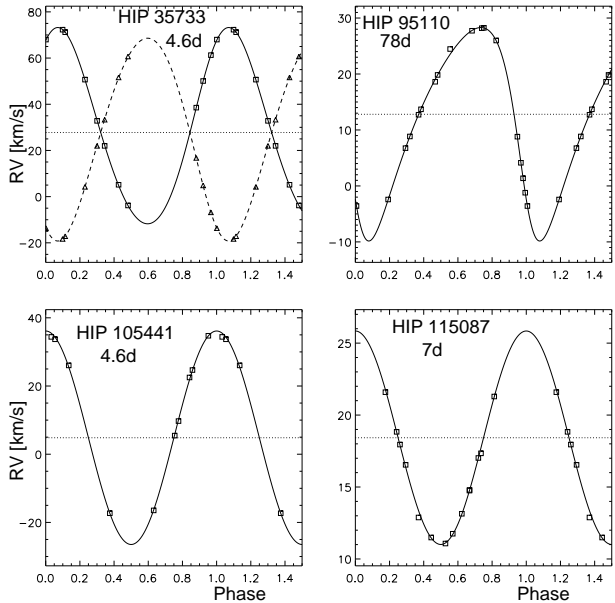}
\caption{RV curves of close binaries.
\label{fig:SB}
}
\end{figure}

Figure~\ref{fig:SB}  gives   the  RV  curves   of  some  spectroscopic
binaries.  Spectroscopic orbital  elements  derived in  this work  are
listed in  Table~\ref{tab:sborb}, while the  visual orbits are  assembled in
Table~\ref{tab:vborb},  in  common   notation.   The  last  column  of
Table~\ref{tab:sborb}   gives   weighted    rms   residuals   to   the
spectroscopic  orbits.   The  combined  orbits are  featured  in  both
tables, duplicating overlapping elements.  In the combined orbits, the
longitude  of   periastron  $\omega_A$  corresponds   to  the  primary
component, and  the position angle  of the visual orbit  $\Omega_A$ is
chosen accordingly  to describe the motion of  the secondary component
in the sky.   Weights are inversely proportional to  the squares of the
measurement  errors  which  are  assigned subjectively  based  on  the
observing  technique   and  adjusted,  if  necessary,   to  match  the
residuals. The  weights of positional measurements and RVs
are balanced,  so that each data  set has $\chi^2/M  \sim 1$. Outliers
are given very low weights by assigning unrealistically large errors.

\begin{deluxetable*}{l l cccc ccc c}    
\tabletypesize{\scriptsize}     
\tablecaption{Spectroscopic orbits
\label{tab:sborb}          }
\tablewidth{0pt}                                   
\tablehead{                                                                     
\colhead{HIP} & 
\colhead{System} & 
\colhead{$P$} & 
\colhead{$T$} & 
\colhead{$e$} & 
\colhead{$\omega_{\rm A}$ } & 
\colhead{$K_1$} & 
\colhead{$K_2$} & 
\colhead{$\gamma$} & 
\colhead{rms$_{1,2}$} \\
& & \colhead{(d)} &
\colhead{(+24\,00000)} & &
\colhead{(deg)} & 
\colhead{(km~s$^{-1}$)} &
\colhead{(km~s$^{-1}$)} &
\colhead{(km~s$^{-1}$)} &
\colhead{(km~s$^{-1}$)} 
}
\startdata
35733  &  Aa,Ab &4.62536& 57407.790  & 0.081 & 330.2    & 42.49 & 43.91 & 27.76   & 0.16 \\
       &  &  $\pm$0.00001 & $\pm$0.010  &  $\pm$0.001 & $\pm$0.8 & $\pm$0.09 & $\pm$0.09 & $\pm$0.04  & 0.14 \\
95110  &  Ba,Bb &78.2379 & 57203.653 & 0.307 & 127.6 & 19.07  & \ldots & 12.80   &  0.13 \\
       &  & $\pm$0.0035 & $\pm$0.090 & $\pm$0.003 & $\pm$0.5 & $\pm$0.09  & \ldots& $\pm$0.04  &  \ldots   \\
105441 & Aa,Ab & 4.6237 & 58026.5495 & 0.000 & 0.0 & 31.30  & \ldots& 4.83   & 0.21 \\
       &       & $\pm$0.0010 & $\pm$0.0069 & fixed & fixed & $\pm$0.14  &\ldots & $\pm$0.17  &  \ldots  \\
105585  &  A,B & 87658   & 55351 & 0.714       & 148.6    & 3.94  & 4.52  & 4.23   &  0.02 \\
        & & fixed  & $\pm$168    &  $\pm$0.029 & $\pm$6.7 & $\pm$0.18  & $\pm$0.20  & $\pm$0.28  &  0.04 \\
105947 &  Aa,Ab &8.7279 & 57899.276  & 0.046           & 117.7    & 27.34  & \ldots & 4.26   & 0.10 \\
       &        & $\pm$0.0003 & $\pm$0.18 & $\pm$0.016 & $\pm$7.7 & $\pm$0.44  & \ldots& $\pm$0.46  &  \ldots \\
105947 &  A,B & 7576   & 57262.8        & 0.363     & 324.4 & 3.34          & 8.66  & $-$0.75   &  \ldots \\
       &        & $\pm$14 & $\pm$9.8    & $\pm$0.003 & $\pm$0.7 & $\pm$0.15  & $\pm$0.14  & $\pm$0.14  & \ldots \\
109951 & A,B   & 18440  & 47849     & 0.452   & 45.2      & 1.71 & \ldots & $-$23.04   & 0.53  \\
       &  & $\pm$1576   & $\pm$60 & $\pm$0.044 & $\pm$9.6 & $\pm$0.17  & \ldots& $\pm$0.13  &  \ldots  \\
109951 &Ba,Bb  & 111.1  &58025.7     & 0.300         & 42.3      & 16.31 & \ldots & $-$22.50    &  0.13  \\
       &  & fixed       & $\pm$2.9   & $\pm$0.078    & $\pm$9.7  & $\pm$1.72 & \ldots& fixed &  \ldots    \\
115087 &  Aa,Ab & 7.8854 & 55456.978 & 0.000 & 0.0   & 7.43  &\ldots & 18.42   & 0.06  \\
       &  & $\pm$0.0001 & $\pm$0.007 & fixed & fixed & $\pm$0.06  & \ldots & $\pm$0.04  &   \ldots 
\enddata 
\end{deluxetable*}

\begin{deluxetable*}{l l cccc ccc}    
\tabletypesize{\scriptsize}     
\tablecaption{Visual orbits
\label{tab:vborb}          }
\tablewidth{0pt}                                   
\tablehead{                                                                     
\colhead{HIP} & 
\colhead{System} & 
\colhead{$P$} & 
\colhead{$T$} & 
\colhead{$e$} & 
\colhead{$a$} & 
\colhead{$\Omega_{\rm A}$ } & 
\colhead{$\omega_{\rm A}$ } & 
\colhead{$i$ }  \\
& & \colhead{(yr)} &
\colhead{(yr)} & &
\colhead{(arcsec)} & 
\colhead{(deg)} & 
\colhead{(deg)} & 
\colhead{(deg)} 
}
\startdata
105585 &   A,B  & 240        & 2010.42         & 0.714      & 0.695             & 279.1    & 148.6       &  88.4  \\
      &          & fixed         & $\pm$0.46    & $\pm$0.029  & $\pm$0.045    & $\pm$1.6   & $\pm$6.7 &  $\pm$2.6\\ 
105947 & A,B   & 20.744     & 2015.656        & 0.363      & 0.1675            & 335.1    & 324.4       &  51.9  \\
       &        & $\pm$0.037    & $\pm$0.027   & $\pm$0.003  & $\pm$0.0005   & $\pm$0.5   & $\pm$0.7 &  $\pm$0.3  \\ 
109951 & A,B   & 50.49      & 1989.88         & 0.451      & 0.2834            & 262.0    &  45.2       &  17.7  \\
       &        & $\pm$4.32     & $\pm$0.16    & $\pm$0.044  & $\pm$0.020    & $\pm$3.0   & $\pm$9.6 &  $\pm$13.6  
\enddata 
\end{deluxetable*}

Table~\ref{tab:rv}  lists individual  RV measurements,  their assigned
errors used for the weighting,  and residuals to the orbits. The first
column  contains  the  {\it   Hipparcos}  number,  the  second  column
identifies the  system. Column (7) specifies  the component (``a''
for the primary and ``b'' for  the secondary), and the last column (8)
indicates the instrument.  The RVs of other visual components measured
here  are  provided   in  Table~\ref{tab:rvconst};  it  includes  some
previously published RVs.  Table~\ref{tab:speckle} lists the positional
measurements used for the calculation of the combined visual/spectroscopic
orbits.

\begin{deluxetable*}{r l c rrr l l}    
\tabletypesize{\scriptsize}     
\tablecaption{Radial velocities and residuals (fragment)
\label{tab:rv}          }
\tablewidth{0pt}                                   
\tablehead{                                                                     
\colhead{HIP} & 
\colhead{System} & 
\colhead{Date} & 
\colhead{$V$} & 
\colhead{$\sigma$} & 
\colhead{(O$-$C)$$ } &
\colhead{Comp.} &
\colhead{Ref.\tablenotemark{a}} \\ 
 & & 
\colhead{(JD $-$2400000)} &
\multicolumn{3}{c}{(km s$^{-1}$)}  &
&
}
\startdata
 35733 & Aa,Ab  &   54782.813 &   $-$3.83 &    0.50  &   0.08 &   a & D \\
 35733 & Aa,Ab  &   54782.813 &   60.67 &    0.50  &   0.17 &   b & D \\
 35733 & Aa,Ab  &   57319.754 &   61.27 &    0.15  &  $-$0.06 &   a & C \\
 95110 & Ba,Bb  &    55447.556 &  24.46 &   0.20   &  0.94  &  a  & F 
\enddata 
\tablenotetext{a}{
C: CHIRON;
D: Du Pont \citep{LCO};
F: FECH \citep{LCO};
L: \citep{Latham2002};
N: NRES.
}
\end{deluxetable*}

\begin{deluxetable}{r l r r l}    
\tabletypesize{\scriptsize}     
\tablecaption{Radial velocities of visual components
\label{tab:rvconst}          }
\tablewidth{0pt}                                   
\tablehead{                                                                     
\colhead{HIP} & 
\colhead{Comp.} & 
\colhead{Date} & 
\colhead{$V$} & 
\colhead{Ref.\tablenotemark{a}} \\ 
 & & 
\colhead{(JD $-$2400000)} &
\colhead {(km s$^{-1}$)}  &
}
\startdata
 35733 & B &   54782.8161 &  28.05 &  D \\  
 35733 & B &   58194.5606 &  28.05 &  C \\
 35733 & B &   58195.4934 &  28.06 &  C \\
 95106 & A &   55447.5389 &  10.35 &  F \\
 95106 & A &   57261.5404 &   8.22 &  C \\
 95106 & A &   57319.5545 &   8.33 &  C \\
 95106 & A &   57983.5757 &   7.06 &  C \\
105569 & C &   55461.6144 &   3.79 &  F \\
105569 & C &   56885.6309 &   3.64 &  C \\
105569 & C &   56895.6894 &   3.66 &  C \\
105569 & C &   57121.9132 &   3.58 &  C \\
105569 & C &   57218.7528 &   3.67 &  C \\ 
105569 & C &   57986.6965 &   3.70 &  C \\
115087 & B &   57986.6855 &   10.93 & C
\enddata 
\tablenotetext{a}{
C: CHIRON;
D: Du Pont;
F: FECH.
}
\end{deluxetable}

\begin{deluxetable*}{r l l rrr rr l}    
\tabletypesize{\scriptsize}     
\tablecaption{Position measurements and residuals (fragment)
\label{tab:speckle}          }
\tablewidth{0pt}                                   
\tablehead{                                                                     
\colhead{HIP} & 
\colhead{System} & 
\colhead{Date} & 
\colhead{$\theta$} & 
\colhead{$\rho$} & 
\colhead{$\sigma$} & 
\colhead{(O$-$C)$_\theta$ } & 
\colhead{(O$-$C)$_\rho$ } &
\colhead{Ref.\tablenotemark{a}} \\
 & & 
\colhead{(yr)} &
\colhead{(\degr)} &
\colhead{(\arcsec)} &
\colhead{(\arcsec)} &
\colhead{(\degr)} &
\colhead{(\arcsec)} &
}
\startdata
105585 & A,B  &    1900.0000 &  286.0 &   1.0000 &    0.100  &   0.7 &   $-$0.048 & M \\
105585 & A,B  &    1928.0000 &  284.0 &   1.1000 &    0.100  &   0.4 &    0.042 & M \\
105585 & A,B  &    2014.7656 &   98.2 &   0.2211 &    0.005  & $-$0.4 &   $-$0.000 & S \\
105585 & A,B  &    2015.7377 &   99.1 &   0.2216 &    0.005  &  0.3   & 0.001  &  S \\
105585 & A,B  &    2017.6823 &   99.2 &   0.2103 &    0.005  &  0.0  & $-$0.001 & S \\
105947 & A,B  &    1991.2500 &  172.0 &   0.1270 &    0.050  & $-$9.3 &  $-$0.014 & H 
\enddata 
\tablenotetext{a}{
H: Hipparcos;
S: speckle interferometry at SOAR;
s: speckle interferometry at other telescopes;
M: visual micrometer measures.
}
\end{deluxetable*}

\section{Individual objects}
\label{sec:obj}

\subsection{HIP 35733}

The visual binary HJ~3957 was  discovered by J.F.W.~Hershel in 1836 at
15\arcsec  ~separation.   Currently the  pair  is  at 7\farcs44.   Its
position is fixed in time  (the discordant discovery measure is likely
in error).  The  components A and B have  matching RVs, proper motions
(PMs), and parallaxes, hence the binary A,B is definitely physical; its
estimated  period is  5\,kyr.  Double  lines in  the component  A were
noted by  \citet{N04}.  RVs  of Aa,  Ab, and B  were measured  once by
\citet{Desidera2006};  their results roughly  agree with  the proposed
orbit,  but are  not used  in the  fit.  One  measure is  published in
\citep{LCO}, the remaining RVs come  from CHIRON. The areas of the two
CCF dips  are similar  and imply  the flux ratio  of 0.90  between the
components Ab  and Aa  of this twin  binary with  the mass ratio  $q =
0.97$.  The 4.6  day period of Aa,Ab is  determined without ambiguity,
while  the small  eccentricity is  significantly different  from zero.
The RV  amplitudes correspond  to the spectroscopic  mass sum  $(M_1 +
M_2) \sin^3 i =  0.305\;{\cal M}_\odot$, while the absolute magnitudes
of  Aa  and Ab  correspond  to  the masses  of  1.22  and 1.19  ${\cal
  M}_\odot$.   Therefore,  the   orbital  inclination  is  $i  \approx
30$\degr.  The width  of both CCF dips implies  rotation velocities of
$\sim$6.5\,km~s$^{-1}$ according  to the calibration  given in Paper~I
\citep{paper1}.     The    synchronous    rotation   corresponds    to
$\sim$12\,km~s$^{-1}$.   Therefore, the  stars Aa  and Ab  seem  to be
aligned and synchronized with  the orbit, considering its inclination.
The mass  of B  estimated from its  absolute magnitude is  1.16 ${\cal
  M}_\odot$,  so all  three stars  in  this system  are similar.   The
RV(B)=28.1\,km~s$^{-1}$  appears constant  according to  its published
measurements in \citep{LCO,Desidera2006}, confirmed with CHIRON; it is
close to the center-of-mass velocity of A.

\subsection{HIP 95106 and 95110}

The two stars HIP~95106 (G0.5V,  $V= 8.16$ mag) and HIP~95110 (K2V, $V
= 10.26$  mag), at  13\farcs66 from each  other, form a  physical wide
binary  HJ~5107 A,B with  an estimated  period of  $\sim$10\,kyr.  The
{\it  Gaia}   DR2  parallax  of   B,  21.49\,mas,  matches   the  HIP2
\citep{HIP2}  parallax of  A,  21.03\,mas.  The  DR2  parallax for  A,
18.82\,mas,  is  slightly  discrepant,   because  it  is  a  0\farcs27
interferometric  binary  TOK~433 Aa,Ab  with  an  estimated period  of
40\,years and estimated masses of  1.0 and 0.70 ${\cal M}_\odot$.  The
RV(A)   of  13.7\,km~s$^{-1}$   was  suspected   to  be   variable  by
\citet{N04};   indeed,   we   note    the   negative   RV   trend   in
Table~\ref{tab:rvconst}, apparently produced  by the orbital motion of
Aa,Ab. This star has been studied in the past as a solar analogue, but
its binarity casts some doubts on the results.

The secondary component B was  identified as a spectroscopic binary in
\citep{survey}; now the 78 day orbital period of Ba,Bb is established.
The estimated mass  of Ba is 0.79 ${\cal  M}_\odot$, hence the minimum
mass of  Bb is 0.41~${\cal M}_\odot$;  its dip is not  detected in the
CCF.  The semimajor  axis of Ba,Bb is 8\,mas.  Very likely, {\it Gaia}
will eventually provide the astrometric orbit of Ba,Bb.

The system HIP 95106/95110 is  thus a 2+2 quadruple. Moreover,   WDS
mentions  it as a  common proper  motion (CPM)  companion   to HIP~94926
(G1V,  WDS~19190$-$3317), at  1649\arcsec ~distance on  the sky.   The  PMs and
parallaxes of these two objects are indeed similar, although not quite
equal.  However,  HIP~94926 has a constant  RV of $-25.7$\,km~s$^{-1}$
according   to  \citet{N04},   which   rules  out   its  relation   to
HIP~95106/95110.

\subsection{HIP 105441}

Like the previous two objects,  this is a wide 26\farcs7 visual binary
HJ~5255  discovered  by  J.~Hershel  in  1835.   The  {\it  Gaia}  DR2
parallaxes  of   the  components  A   and  B,  31.27  and   31.71  mas
respectively, as  well as their matching  PMs and RVs,  leave no doubt
that  this binary  is physical.   The period  of A,B  is  estimated as
$\sim$20~kyr.   According to  \citet{Shkolnik2017} and  to  some other
authors, the system is a  member of the $\beta$~Pictoris moving group;
lithium lines  were found  in the spectrum  of B (spectral  type K7V).
The component A,  of K2V spectral type, is  a chromospherically active
star V390~Pav.   \citet{N04} identified it as  a spectroscopic binary,
and it  was treated  as such in  several papers, although  the orbital
period remained unknown.   \citet{Messina2017} determined the rotation
period of 5.50$\pm$0.02 days. It is  close to the binary period of 4.6
days  found here.   The component  Aa thus  rotates sub-synchronously.
The  width  of   its  CCF  dip  corresponds  to   $V  \sin  i  \approx
7.5$\,km~s$^{-1}$.  The  estimated mass of  Aa, 0.81~${\cal M}_\odot$,
implies the minimum mass of Ab of 0.26~${\cal M}_\odot$; its lines are
not detected in the CCF. The orbit of Aa,Ab is circular.

\begin{figure}
\plotone{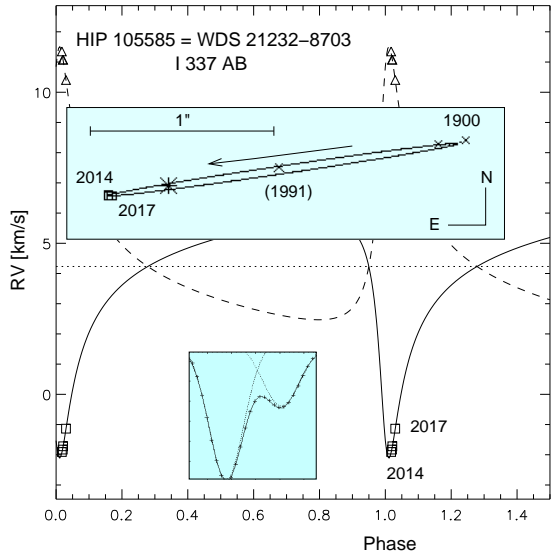}
\caption{Visual  and   spectroscopic  orbit  of   HIP~105585  A,B  (WDS
  21232$-$8703, I~337). The RV curve with  coverage from 2014 to 2017
  is shown. The upper insert shows the orbit in the sky, the lower
  insert is the CCF with two fitted Gaussians. 
\label{fig:I337}
}
\end{figure}

\begin{figure*}
\plotone{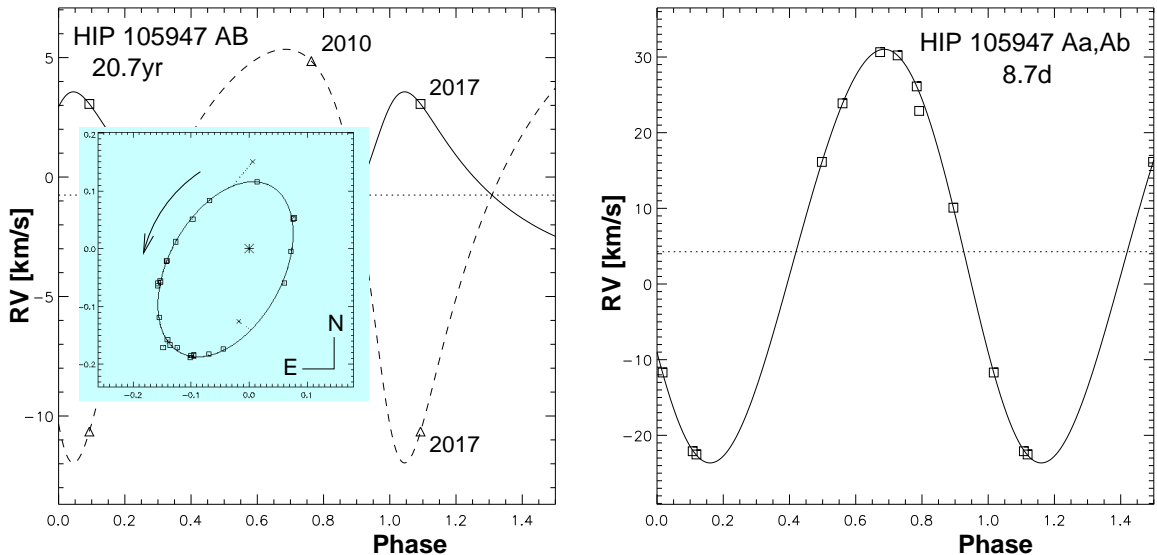}
\caption{The triple system HIP~105947. The orbit of the outer binary A,B
with $P = 20.7$ yr  is shown on the left (the insert displays the
positions on the sky). The right-hand plot shows the RV curve of the
subsystem Aa,Ab with $P=8.7$ days.
\label{fig:105947}
}
\end{figure*}

\begin{figure*}
\plotone{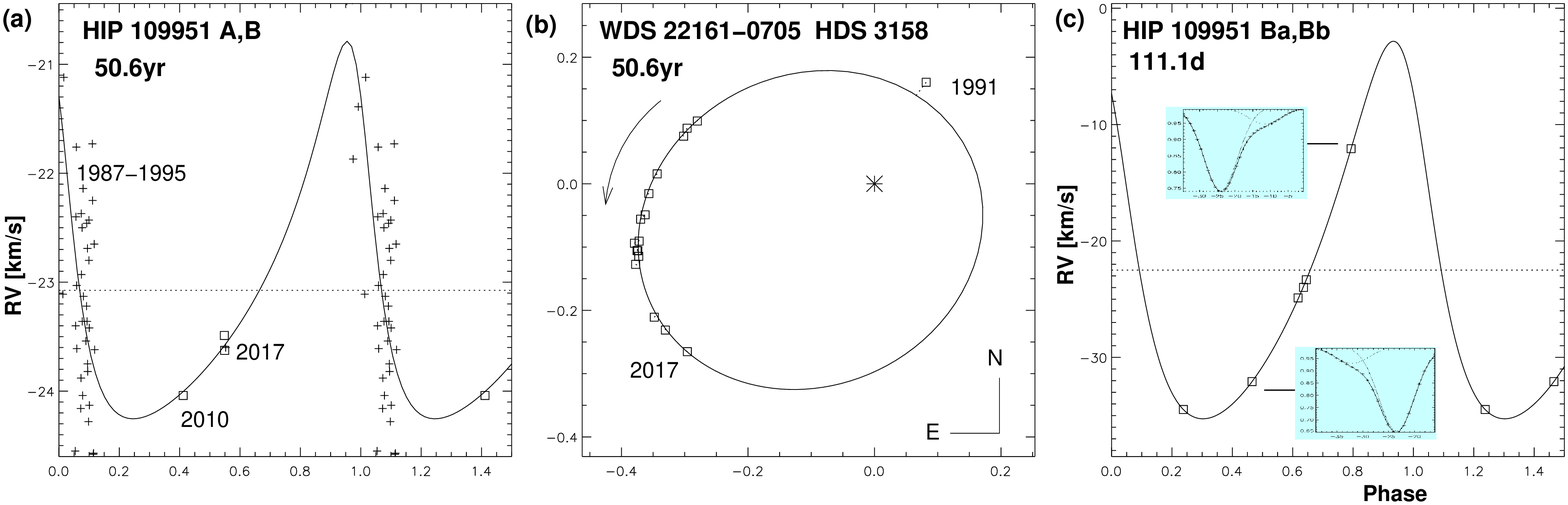}
\caption{The triple system HIP~109951. For the  outer binary A,B with $P =
  50.6$ yr, the panel (a) shows the RV curve, where crosses denote
  the RVs from \citet{Latham2002} and squares are the RVs measured at
  CTIO. The middle panel(b) is the orbit in the plane of the sky. The
  right panel (c) shows the preliminary orbit of the subsystem Ba,Bb with $P=111$ days;
  the two inserts are asymmetric CCFs recorded in 2010 and 2017 and fitted
  by double Gaussians.   
\label{fig:109951}
}
\end{figure*}

\subsection{HIP 105585 and 105569}

This  visual  triple system  is  located  at  3\degr ~from  the  South
celestial  pole.  The  17\farcs9  outer pair  AB,C  was discovered  by
J.~Hershel (HJ 5192).  The component C (HIP~105569) has the {\it Gaia}
DR2  parallax of  13.99\,mas,  adopted  here as  the  distance to  the
system;  The DR2 gives no parallax for A, while the HIP2 parallax
of A, 30.24$\pm$4.85\,mas, is  obviously discrepant. Many binaries with
separations  around 20\arcsec  ~have problematic  {\it  Hiparcos} data
caused by the design of its measurement system.  The components AB and
C are  located on the main  sequence, meaning that  the G8III spectral
type  of AB  given by  SIMBAD is  incorrect. The  small  PM difference
between AB and C is caused by the orbital motion of AB.

The  inner  pair  A,B  (I~337)  has  closed  down  from  the  1\arcsec
~separation in 1900 to 0\farcs2 at present.  Its double-lined spectrum
\citep{survey}     suggested    presence     of     a    spectroscopic
subsystem. However,  further monitoring revealed that the  RVs of both
components  are  remarkably  stable  in  time  and  match  the  visual
components A  and B.   The binary is  actually near periastron  of its
eccentric  long-period orbit.   Such preliminary  orbit with  $P= 240$
years is computed here by  combining RVs with a few available position
measurements.  Proximity to the  pole makes the correction of position
angles for  precession essential in the orbit  calculation.  The orbit
is  seen almost exactly  edge-on (inclination  88\fdg4).  In  1991 the
separation  of A,B  was around  0\farcs4; its  non-resolution  by {\it
  Hipparcos} is  possibly related to the mentioned  above problem with
this object. 

Spectroscopic orbits with century-long periods are unusual.  The Ninth
Catalog  \citep{SB9} contains  only two  entries with  longer periods:
HIP~10952 with $P=319.4$ years  and the Cepheid T~Mon (HIP~30541) with
$P=257$ years.

The ratio of the CCF dip areas corresponds to the magnitude difference
between  A and  B of  $\Delta V  = 0.86$  mag;  speckle interferometry
measured  $\Delta I  =  0.84 \pm  0.05$ mag.  The  masses of  A and  B
estimated  from  their absolute  magnitudes  are  1.0 and  0.86~${\cal
  M}_\odot$. They  match the spectroscopic  mass ratio $q =  0.87$ and
the masses derived  from the combined orbit. The  orbit corresponds to
the parallax of 14.67\,mas.

The  RV   of  the  visual   tertiary  component  C  was   measured  at
3.30\,km~s$^{-1}$  by  \citet{N04}. Five  measurements  of RV(C)  with
CHIRON average at 3.65$\pm$0.02\,km~s$^{-1}$. As far as we can tell, C
is a 1.0~${\cal M}_\odot$ single star.

\subsection{HIP 105947}

This  close 0\farcs1  binary was  first resolved  by  {\it Hipparcos}.
Follow-up  observations  with  speckle  interferometry  revealed  fast
orbital motion, leading to the calculation of the orbit with a 20 year
period.  It was  revised for the last time  by \citet{Tok2015c} and is
slightly corrected here.  The improved coverage of this orbit leads to
very accurate  elements ($P = 20.74  \pm 0.04$ yr). 

Double  lines,  noted  by  \citet{N04} and  \citet{survey},  indicated
presence  of  a spectroscopic  subsystem.   The  RV  of the  strongest
component Aa varies fast, while the weaker component is stationary and
corresponds to  the visual secondary  B. Monitoring of this  object in
2017 with CHIRON  and NRES results in the  8.7 day spectroscopic orbit
of Aa,Ab.  This orbit is nearly, but not quite, circular. Both dips in
the CCF are narrow and correspond  to a slow axial rotation $V \sin i$
of $\sim$2.5~km~s$^{-1}$.

The RVs  of the component B  measured in 2010 and  2017 are different,
reflecting its  motion in the outer  orbit. These RVs, as  well as the
center-of-mass  velocity of  Aa,Ab  in 2017,  serve  to establish  the
spectroscopic   elements  of  the   outer  orbit,   complementing  the
positional measures  and identifying  the correct ascending  node. The
single measure of the RV(Aa) in 2010 is not useful for the outer orbit
because its  interpretation depends  on the exact  value of  the inner
period. As  the three  spectroscopic elements of  the outer  orbit are
derived  from  three  measurements,  the  RV residuals  are  zero.   I
attempted to fit all data  by two orbits simultaneously using the {\tt
  orbit3.pro} code \citep{TL2017}, but  the results do not differ from
the separate solutions. The ``wobble''  in the motion of A,B caused by
the   8--day   subsystem   is   $\sim$0.4\,mas,  too   small   to   be
detectable. However, it should be seen by {\it Gaia}.

Differential speckle  photometry of A,B from various  sources leads to
$\Delta V = 1.47$ mag, hence the individual $V$ magnitudes of 7.77 and
9.24  for   Aa  and  B  and   the  masses  of   1.25  and  0.96~${\cal
  M}_\odot$. The minimum mass of  Ab is 0.37~${\cal M}_\odot$, and the
total estimated mass sum  is 2.57~${\cal M}_\odot$.  Together with the
visual orbit, it leads to the dynamic parallax of 16.23\,mas, while the
{\it  Gaia}  DR2  parallax  is  15.36\,mas.   Independently  of  the
parallax, the  RV amplitudes  and the inclination  of the  outer orbit
correspond to the  masses of 1.61 and 0.62~${\cal  M}_\odot$ for A and
B, respectively.  All data match reasonably well.

\subsection{HIP 109951}

This is a  slightly metal-deficient G5V star with  a fast PM, G27--21.
The binary  system A,B was  discovered by {\it  Hipparcos} (HDS~3158),
and  its first  visual orbit  with  $P =  80.6$ year  was computed  by
\citet{Cvetkovic2014}.   \citet{Latham2002}  found   that  the  RV  is
changing  slowly,  presumably  owing  to  the  motion  in  the  visual
orbit. However, \citet{survey} recorded in 2010 an asymmetric CCF. The
RV  difference between  two  components of  this  blended profile  was
larger than expected from the visual orbit, suggesting the presence of
an inner spectroscopic subsystem. Two  more asymmetric CCFs were recorded
in 2017, confirming the existence of the subsystem.
 
Differential speckle photometry of the  resolved components A and B by
several groups  yields consistent results: $\Delta V_{\rm  AB} = 1.87$
mag and $\Delta I_{\rm AB} = 1.39$ mag. The individual $V$ magnitudes,
8.92 and 10.79 mag, correspond  to the main sequence stars with masses
1.00 and 0.77 ${\cal M}_\odot$. The ratio of the areas of the CCF
dips is 0.2, matching the flux ratio of A and B, so the secondary dip
corresponds to the visual component B.  Its RV is variable because of
the subsystem Ba,Bb, while the RV of the main dip varies slowly,
following the orbit of A,B. The RVs measured by \citet{Latham2002}
correspond to the blended lines of A and Ba and are affected by both
orbits. 

I determined  the combined orbit  of A,B using all  available position
measures and  the RVs of component  A (Figure~\ref{fig:109951}).  Half
of  this  orbit  is  now  covered. The  RVs  from  \citet{Latham2002},
measured in the  period from 1987.5 to 1994.7,  have the rms residuals
of  0.82\,km~s$^{-1}$. The  residuals  to the  outer  orbit contain  a
periodic  component  with   $P  =  111$  days  and   an  amplitude  of
1.0\,km~s$^{-1}$.  When this  signal is  subtracted,  the residuals
decrease  to  0.47\,km~s$^{-1}$, and  show  no  systematic trend  with
time. The periodic  signal is caused by blending  of the spectra of
stars A and Ba, and the period corresponds to the subsystem Ba,Bb.

Having established  the period of  the subsystem, I derived  its 
preliminary orbit from  the CTIO data.  The RV of  Ba measured in 2010
was corrected by $-0.55$\,km~s$^{-1}$  to account for the outer orbit.
The RVs of A derived  from three unresolved (blended) CCFs (JD 2457983
-- 986) show a positive trend  over 3 days, with a slightly decreasing
width of the CCF and  a slightly increasing contrast. The residuals of
these RVs to the outer orbit  are interpreted as a result of blending.
Approximate RVs  of Ba on those  three nights are derived  using the known
ratio of  the CCF  areas of  A and Ba.   The orbit  of Ba,Bb  shown in
Figure~\ref{fig:109951} (c) is still tentative.  It can be improved by
planning  further  observations  at  phases  where the  two  dips  are
separated.   The  phase of  the  periodic  component  detected in  the
residuals of the  published RVs matches the new  orbit, confirming its
period. 

The minimum  mass of  Bb is 0.26\,${\cal  M}_\odot$. Its sum  with the
estimated masses of  A and Ba is 2.03\,${\cal  M}_\odot$ and, together
with the  visual orbit, implies the dynamical  parallax of 15.85\,mas.
The visual  orbit and the  {\it Gaia} DR2  parallax lead to  the total
mass  sum  of 2.34\,${\cal  M}_\odot$,  in  agreement  with the  above
estimates.  The  mass of B derived from  the RV amplitude of  A in the
outer orbit  and its inclination  is 0.99\,${\cal M}_\odot$,  close to
the  estimated  1.02\,${\cal M}_\odot$.   However,  the  error of  the
inclination is relatively large, hence the ``spectroscopic'' mass of B
is poorly constrained.

The semimajor axis of the Ba,Bb orbit is 7\,mas, so a 1.7 mas ``wobble''
in  the positions of  A,B is  expected.  In  principle, the  effect is
detectable using speckle astrometry  of A,B, although its amplitude is
comparable to  the measurement  errors. So far, no attempt  has been  made to
measure the  wobble, although the  tools for doing57986.6855  10.933 this  are available
\citep{TL2017}.  The weighted rms residuals from the current orbit are
3\,mas in both coordinates.  

The   object  belongs   to   one  of   the   {\it  Kepler-2}   fields.
\cite{Lund2016} determined  stellar parameters from  the seismological
analysis, assuming a  single star.  They give two  possible values for
the mass,  0.82~${\cal M}_\odot$ or 0.93~${\cal M}_\odot$;  the age is
about 12\,Gyr. 

Accurate  astrometry  with  {\it   Gaia}  will  eventually  bring  new
information on this triple  system. The astrometric signal should show
an acceleration caused by the 50 year  orbit and the wobble with a period
of 111 days.  The orbital parameters determined here  will help in the
{\it Gaia} data reduction.

\subsection{HIP 115087}

This  is a  young,  chromospherically  active G1V  star  in the  solar
neighborhood (GJ~9819).  Like other objects in this program, this is a
wide 45\farcs2  binary HJ~5392 discovered by  J.~Hershel. However, the
component B of  spectral type K0III has different PM   and RV, so
it is  optical (unrelated).  The variable  RV of the  main component A
was detected by \citet{N04}  and confirmed by \citet{survey}, where an
orbital period of 7.88 days was suggested.  Five new observations with
NRES  and  four with CHIRON confirmed this  period.  The minimum
mass  of  the  secondary   component  is  quite  small,  0.073\,${\cal
  M}_\odot$.  

This is a rare binary with a short period and a low-mass companion, in
the  so-called  brown dwarf  desert  regime.   However,  the small  RV
amplitude  can  result  from  small  orbital  inclination,  while  the
secondary component has  a stellar mass.  The  four CCFs measured
  with CHIRON are narrow and correspond  to $V \sin i = 2.55 \pm 0.01$
  km~s$^{-1}$    according   to    the   calibration    presented   in
  \citep{paper1}. The  primary star  of one solar  radius synchronized
  with the orbit  has an equatorial speed of  6.35 km~s$^{-1}$, so the
  likely inclination  of the  rotation axis and  the orbit is $i  \approx
  24\degr$. The inferred mass of the secondary component is then 
  0.2\,${\cal M}_\odot$.

\section{Summary}
\label{sec:sum}

This  work is  a small  contribution  to improving  the statistics  of
hierarchical  multiplicity  in  the  solar  neighborhood.   Previously
unknown spectroscopic  orbits of six inner  subsystems are determined,
as well as several outer orbits.   Addition of these orbits to the
  $\sim$500 known  inner spectroscopic orbits  in hierarchical systems
  within  100  pc \citep{MSC}  does  not  justify  a full  statistical
  re-analysis.     The  difficulty  of such  a  study involving  time
scales from days to centuries is obvious.

Orbits of most late-type dwarfs with periods below 10 days are tidally
circularized. However,  among three such subsystems  studied here, two
(HIP  35733 Aa,Ab  and HIP  105947 Aa,Ab)  have small  but significant
eccentricities.   The outer  orbit  of  HIP~35733 is  too  wide for  a
dynamical interaction with the  inner 4.6--day subsystem. As suggested
by \citet{Moe2018},  the triple-star  dynamics could produce  an inner
binary with  a much longer period  and a high  eccentricity. The inner
orbit then slowly becomes circular  owing to equilibrium tides; we
may be witnessing the last stage of circularization in these systems.

The  results will  be relevant  for interpretation  of the  {\it Gaia}
astrometry  and, eventually,  for accurate measurement of stellar
  masses and  other parameters. This  mission is limited both  by its
time span  (5 years) and by  the observing cadence.   Despite the high
precision  that  gives access  to  astrometric  orbits  of even  close
binaries, it  will be  difficult to derive  such orbits  without prior
knowledge  of  their  periods.  Therefore,  ground-based  spectroscopy
appears to be  an essential complement to the  {\it Gaia} mission. Its
own spectroscopic  capability is much  inferior to the  resolution and
accuracy  of  spectrometers   like  CHIRON.   Likewise,  ground--based
speckle monitoring  of visual binaries complements {\it  Gaia} on long
time scales.




\acknowledgements

I  thank   the  operators  of   the  1.5-m  telescope   for  executing
observations  of  this  program  and  the  SMARTS  team  at  Yale  for
scheduling and pipeline processing.  Re-opening of CHIRON in 2017 was
largely due to the enthusiasm and energy of T.~Henry. Access to the
newly commissioned NRES spectrometer has been important for this
project; I thank N.~Volgenau and T.~Brown from the Las Cumbres
observatory for  help with this instrument and its data products.

This work  used the  SIMBAD service operated  by Centre  des Donn\'ees
Stellaires  (Strasbourg, France),  bibliographic  references from  the
Astrophysics Data  System maintained  by SAO/NASA, and  the Washington
Double Star Catalog maintained at USNO.

{\it Facilities:}  \facility{CTIO:1.5m}, \facility{SOAR}, \facility{LCO:NRES} 










\end{document}